\documentclass[rmp,aps]{revtex4-1}
\usepackage{amsfonts}
\usepackage{amssymb}
\usepackage{amsmath,amsthm}
\usepackage{graphicx}
\usepackage{setspace}
\usepackage{CJK}
\usepackage{moresize}
\usepackage[utf8]{inputenc}
\usepackage[english]{babel}
\usepackage{xcolor}

\usepackage{booktabs}
\usepackage{siunitx}
\usepackage{booktabs, makecell, multirow}

\newcommand{\begM}{\begin{multline}}
\newcommand{\eM}{\end{multline}}

\def\b{\beta}
\def\a{\alpha}
\newcommand{\m}{\mu}
\newcommand{\n}{\nu}
\newcommand{\p}{\partial}

\def\){\Big)}
\def\({\Big(}

\def\r{\rho}

\def\k{\kappa}
\def\t{\tau}
\def\g{\gamma}
\def\x{\chi}

\begin{document}

\title[Curvature tensors of higher-spin gauge theories derived from general Lagrangians]{Curvature tensors of higher-spin gauge theories derived from general Lagrangian densities}

\author{Mark Robert Baker}
\author{Julia Bruce-Robertson}

\affiliation{Department of Physics and Astronomy, Western University, London, ON, Canada N6A 3K7}

\affiliation{The Rotman Institute of Philosophy, Western University, ON, Canada N6A 5B7}

\email{Electronic address: mbaker66@uwo.ca}
\email{Electronic address: jbrucero@uwo.ca}

\begin{abstract}
Curvature tensors of higher-spin gauge theories have been known for some time. In the past, they were postulated using a generalization of the symmetry properties of the Riemann tensor (curl on each index of a totally symmetric rank-$n$ field for each spin-$n$). For this reason they are sometimes referred to as the generalized 'Riemann' tensors. In this article, a method for deriving these curvature tensors from first principles is presented; the derivation is completed without any a priori knowledge of the existence of the Riemann tensors or the curvature tensors of higher-spin gauge theories. To perform this derivation, a recently developed procedure for deriving exactly gauge invariant Lagrangian densities from quadratic combinations of $N$ order of derivatives and $M$ rank of tensor potential is applied to the $N = M = n$ case under the spin-$n$ gauge transformations. This procedure uniquely yields the Lagrangian for classical electrodynamics in the $N = M = 1$ case and the Lagrangian for higher derivative gravity (`Riemann' and `Ricci' squared terms) in the $N = M = 2$ case. It is proven here by direct calculation for the $N = M = 3$ case that the unique solution to this procedure is the spin-3 curvature tensor and its contractions. The spin-4 curvature tensor is also uniquely derived for the $N = M = 4$ case. In other words, it is proven here that, for the most general linear combination of scalars built from $N$ derivatives and $M$ rank of tensor potential, up to $N=M=4$, there exists a unique solution to the resulting system of linear equations as the contracted spin-$n$ curvature tensors. Conjectures regarding the solutions to the higher spin-$n$ $N = M = n$ are discussed.
\end{abstract}

\maketitle

\tableofcontents

\section{Motivation}

\linespread{1}
\normalsize

Higher-spin gauge theories describing free massless fields are well established in the literature. These theories have gauge transformations and curvature tensors that have been generalized for any spin-$n$ model considered \cite{fronsdal1978,fang1978,deWit1980,damour1987,sorokin2005}. In the past the curvature tensors were postulated based on symmetry properties of the Riemann tensors (by taking the curl on each index of a totally symmetric rank-$n$ field for each spin-$n$ \cite{damour1987}). Here we present a method to derive these curvature tensors from first principles; they are derived by direct calculation without any knowledge of the existence of the Riemann tensors or curvature tensors of higher-spin gauge theories.

The higher-spin curvature tensors, sometimes referred to as the generalized `Riemann' curvature tensors for their generalization as $n$ pairs of antisymmetric indices for each spin-$n$ model analogous to the Riemann tensor in the $n = 2$ case, are independently gauge invariant under the spin-$n$ gauge transformations. They are of particular interest in the Maxwell-like higher spin models that consider equations built from the divergence of these curvature tensors, which are analogous to Maxwell's equation in the spin-1 case. The Maxwell-like higher spin models have been primarily worked out in a series of papers by Francia et al. \textcite{francia2012,francia2014,bekaert2015,francia2017}. The curvature tensors also allow for the generalization of the dual formulation of higher spin models used commonly for models built with the $n=1$ field strength tensor $F^{\mu\nu}$ and $n =2$ Riemann tensor $R^{\mu\nu\alpha\beta}$ \cite{henneaux2016b,danehkar2019}. Generalization of the curvature tensors can be found in \cite{sorokin2005,bekaert2006}. In the past these generalizations have been developed by extrapolating from the symmetries of lower spin-$n$ models, rather than by derivation from some general principles. The latter approach is what we will develop in this article: for each spin-$n$ model, we will independently and uniquely derive the curvature tensors and their contractions (the 'Ricci' forms of the curvature tensors) from a general linear system of scalars, without any a priori knowledge of their existence.
No knowledge of the curvature tensors or required symmetries is necessary for this procedure; only the form of the spin-$n$ gauge transformations given in equations (\ref{spin1trans}) to (\ref{spin4trans}) is needed to perform this derivation.

Recent research developed a procedure for deriving completely gauge invariant Lagrangians by considering general linear systems of scalars under a particular gauge transformation \cite{baker2019}. The general Lagrangian density is expressed in terms of free coefficients which are solved for such that the resulting Lagrangian density is exactly gauge invariant (not only invariant up to a surface term). The scalars are built from quadratic combinations of $N$ order of derivatives of $M$ rank of tensor potentials. When this procedure is applied to the $N = M = 1$ case and the spin-1 gauge transformation $A_\mu' = A_\mu + \partial_\mu \xi$ is used, the Lagrangian $\mathcal{L} = C F_{\mu\nu} F^{\mu\nu}$ is uniquely derived. When it is applied to the $N = M = 2$ case and a spin-2 gauge transformation (linearized diffeomorphism) $h_{\mu\nu}' = h_{\mu\nu} + \partial_\mu \xi_\nu + \partial_\nu \xi_\mu$ is used, the Lagrangian $\mathcal{L} = \tilde{a} R_{\mu\nu\alpha\beta} R^{\mu\nu\alpha\beta} + \tilde{b} R_{\mu\nu} R^{\mu\nu} + \tilde{c} R^2$ is uniquely derived.

The natural question that arose was what would occur if this procedure were applied to the $N = M = n$ case. Since the $N = M = 1$ case yields the scalar $F_{\mu\nu} F^{\mu\nu}$, built from the field strength tensor of electrodynamics, and the $N = M = 2$ yields the scalars $R_{\mu\nu\alpha\beta} R^{\mu\nu\alpha\beta}$, $R_{\mu\nu} R^{\mu\nu}$ and $R^2$, built from the linearized Riemann tensor and its contractions, it was conjectured that to derive an exactly gauge invariant Lagrangian from this procedure, it would be necessary to have contraction of independently gauge invariant `field strength' (curvature) tensors. The validity of this conjecture is further investigated in this article.

To explore the extension to $N = M = n$, we started with $N = M = 3$ under the spin-3 gauge transformation $\phi_{\a \b \r}' = \phi_{\a \b \r} + \p_{\a} \lambda_{\b \r} + \p_{\b} \lambda_{\a \r} + \p_{\r} \lambda_{\a \b}$, where $\lambda_{\b \r}$ is a symmetric gauge parameter. As in \cite{baker2019}, we will only consider totally symmetric fields $\phi$ in this article, however in principle this procedure should work for any field symmetries, such as antisymmetric field models \cite{mckeon2004}. In this article we will show that, as it did for spin-1 and spin-2 in \cite{baker2019}, the procedure yields a solution of several scalar invariants, which turn out to be the contracted spin-3 curvature tensor, $K^{\t \n \k \m \x \g}$ \cite{bekaert2006} (sometimes referred to as the `Riemann' tensor generalization), and its `Ricci' tensors, $K^{\t \n \k \m }$ and $K^{\t \n }$. In other words the higher spin curvature tensors can be derived from this procedure without a priori knowledge of their existence.
 Extending this to $N = M = 4$ again yields the contraction of the independently gauge invariant curvature tensor $K^{\a \b \g \x \m \k \n \t}$, namely the spin-4 generalization of the `Riemann' curvature tensor. For $N = M = 5$ and greater, the calculations became too difficult for us to do by hand (since each had thousands of scalar terms in the general expression), so, instead, conjectures about the nature of the spin-$n$ Lagrangians, based on the cases $n=1,2,3,4$, are made at the end of the article. 
 We note that higher spin models \cite{mckeon2006,maloomeh2008} and the Lagrangian formulation for higher spin models is well researched from various points of view \cite{deMedeiros2003,fotopoulos2009,buchbinder2012}, but primarily these consider conventional (second order) spin-$n$ equations of motion. Here, when we consider $N = M = n$ models, we have terms in the Lagrangian that are quadratic combinations of $n$ order of derivatives and $n$ rank of tensor potential. These types of higher spin Lagrangians are less developed in the literature \cite{francia2010}. For our purposes, we use these Lagrangians to derive the curvature tensors of higher-spin gauge theories without any a priori knowledge of their existence. 

We acknowledge that there are several issues related to the unitary and renormalizability of higher derivative theories, that continue to be worked out in the literature \cite{abe2019,brandt2020}. At no point do we consider the higher derivative Lagrangian densities we write down for higher $N = M = n$ cases to avoid or solve these problems: our motivation is purely to give a derivation of the well known curvature tensors of higher spin gauge theories without a priori knowledge of their existence, or of existence of the Riemannian tensors. We do this because previously they have been merely postulated using a generalization of the symmetry properties of the Riemann tensor (curl on each index of a totally symmetric rank-n field for each spin-$n$). Our method more naturally and independently obtains them alongside the Riemannian tensors and electrodynamics field strength tensor, since these tensors are the natural outcomes for the $N = M = 2$ and $N = M = 1$ derivations, respectively. The models associated to the Lagrangian densities we use to derive the higher spin curvature tensors have no, to our knowledge, new predictive insight of physical phenomena.

The article will be structured as follows. First, we will detail the procedure for deriving completely gauge invariant models by considering the spin-1 case and discuss generalizations of the gauge transformations to spin-$n$. Next, we will start from the general Lagrangian for the $N = M = 3$ case and, under the spin-3 gauge transformation, show how this yields precisely the contractions of the `Riemann' and `Ricci' curvature tensors of spin-3. This process will then be repeated for spin-4. Finally, we will provide conjectures about the behaviour of Lagrangians derived from the procedure for $N = M = n$, giving some indication of how these Lagrangians will be built for the spin-$n$ case.

\section{Derivation of the curvature tensors of higher-spin gauge theories}

In \cite{baker2019}, a procedure for deriving exactly gauge invariant Lagrangians is outlined in detail for the case $N = M = 1$ and $N = M = 2$. This procedure involves writing down the most general scalars for each case and solving for free coefficients such that the resulting Lagrangian density is exactly invariant under the gauge transformation being considered. 
For the $N = M = 1$ case, for which the most general scalar is the sum of all possible scalars of the form $\partial A \partial A$, when the spin-1 gauge transformation $A_\mu' = A_\mu + \partial_\mu \xi$ is applied, the resulting Lagrangian is $\mathcal{L} = C F_{\mu\nu} F^{\mu\nu}$. 
For the $N = M = 2$ case, for which the most general scalar is the sum of all possible scalars of the form $\partial \partial h \partial \partial h$, when the spin-2 gauge transformation $h_{\mu\nu}' = h_{\mu\nu} + \partial_\mu \xi_\nu + \partial_\nu \xi_\mu$ is applied, the resulting Lagrangian decouples into three independently gauge invariant scalars that turn out to be the linearized 'Riemann' and `Ricci' tensors; $\mathcal{L} = \tilde{a} R_{\mu\nu\alpha\beta} R^{\mu\nu\alpha\beta} + \tilde{b} R_{\mu\nu} R^{\mu\nu} + \tilde{c} R^2$.
The obvious next step is to generalize the procedure for the $N = M = n$ case. This generalization, as we will show, can be used to derive the curvature tensors of higher-spin gauge theories without any a priori knowledge of their existence. Since the generalization of the scalars is fixed by the procedure, the only required input is the gauge transformation which we require the models to be invariant under. For this, we require the spin-$n$ gauge transformations that are adopted in the literature \cite{sorokin2005,bekaert2006}. These generalizations are of the form

\begin{equation}
A_\mu' = A_\mu + \partial_\mu \xi  \label{spin1trans}
\end{equation}
\begin{equation}
h_{\mu\nu}' = h_{\mu\nu} + \partial_\mu \xi_\nu + \partial_\nu \xi_\mu  \label{spin2trans}
\end{equation}
\begin{equation}
\phi_{\a \b \r}' = \phi_{\a \b \r} + \p_{\a} \lambda_{\b \r} + \p_{\b} \lambda_{\a \r} + \p_{\r} \lambda_{\a \b}  \label{spin3trans}
\end{equation}
\begin{equation}
\phi_{ \b \x \k \t}'  = \phi_{ \b \x \k \t} +  \p_{\b} \lambda_{ \x \k \t} + \p_{\x }\lambda_{\b \k \t} + \p_{ \k}\lambda_{\b \x  \t} + \p_{\t} \lambda_{\b \x \k }  \label{spin4trans}
\end{equation}

where the potentials, $\phi$, and the gauge parameters, $\lambda$, are completely symmetric in all indices; this generalization continues to all $n$. We now have everything we need to apply the procedure for the $N = M = n$ case.

\subsection{\boldmath Scalars built from contracted spin-$n$ curvature tensors for $N = M= 3$}

For the $N = M = 3$ case, the most general Lagrangian density is the sum of all possible unique scalars of the form $\p \p \p \phi \p \p \p \phi$. 
This set of unique scalars is obtained by considering all the possible summation patterns that could occur in a scalar of the form $\p \p \p \phi \p \p \p \phi$. First, recognize that each scalar is the contraction of two terms, $\p \p \p \phi$ and $\p \p \p \phi$. Let us call these terms $A$ and $B$. We can then group the possible scalars into four categories: scalars in which $A$ and $B$ each have 6 free indices (i.e. all the summation occurs between $A$ and $B$, not within either), scalars in which $A$ and $B$ each have 4 free indices (i.e. summation occurs between $A$ and $B$ as well as within each), scalars in which $A$ and $B$ each have 2 free indices, and scalars in which $A$ and $B$ each have no free indices. Now, within each category, we consider all the possible ways the indices can sum. It is possible that an index on a derivative in $A$ sums with an index on another derivative in $A$, with an index on $\phi$ in $A$, with an index on a derivative in $B$, or with an index on $\phi$ in $B$. Since $\phi$ is symmetric, these are the only unique possibilities. Likewise, it is possible that an index on $\phi$ in $A$ sums with another index on $\phi$ in $A$, with an index on a derivative in $A$, with an index on $\phi$ in $B$, or with an index on a derivative in $B$. 

Using these possibilities, we form all possible combinations of index sums within the contraction such that the resulting $A$ and $B$ terms have the given number of free indices. By writing out a term for each possible summation pattern within each category, we obtain a comprehensive set of all the unique scalars of the form $\p \p \p \phi \p \p \p \phi$. This set leads to the most general Lagrangian density

\begin{equation}
\begin{split}
\mathcal{L} {}&  =
C_{1} \p_{\x} \p_{\t} \p_{\k} \phi_{\g \m \n} \p^{\x} \p^{\t} \p^{\k} \phi^{\g \m \n} 
+ C_{2} \p_{\x} \p_{\t} \p_{\k} \phi_{\g \m \n} \p^{\x} \p^{\t} \p^{\g} \phi^{\k \m \n} 
+ C_{3} \p_{\x} \p_{\t} \p_{\k} \phi_{\g \m \n} \p^{\x} \p^{\m} \p^{\n} \phi^{\k \t \g} 
\\
&{}+ C_{4} \p_{\x} \p_{\t} \p_{\k} \phi_{\g \m \n} \p^{\g} \p^{\m} \p^{\n} \phi^{\x \t \k}
+ C_{5}  \p_{\x} \p_{\t } \p_{\k} \phi^{\x \m \n} \p_{\g} \p^{\k} \p^{\t} \phi^{\g}_{\m \n} 
+ C_{6}  \p_{\x} \p_{\t } \p_{\k} \phi^{\x \m \n} \p_{\m} \p_{\n} \p_{\g} \phi^{\g \k \t} 
\\
{}&+ C_{7}  \p_{\x} \p_{\t } \p_{\k} \phi^{\x \m \n}  \p^{\k} \p_{\m} \p_{\g} \phi^{\g \t}_{ \n} 
+ C_{8}  \p_{\x} \p_{\t } \p_{\k} \phi^{\x \m \n} \p_{\g} \p^{\g} \p^{\k} \phi^{\t}_{\m \n}
+ C_{9}  \p_{\x} \p_{\t } \p_{\k} \phi^{\x \m \n} \p_{\g} \p^{\g} \p_{\n} \phi^{\k \t }_{\m}
\\
{}&+ C_{10}  \p_{\x} \p_{\t } \p_{\k} \phi^{\x \m \n} \p^{\t} \p^{\k} \p_{\m} \phi^{\g}_{\g \n}
+ C_{11} \p_{\x} \p_{\t } \p_{\k} \phi^{\x \m \n} \p^{\t} \p_{\m} \p_{\n} \phi^{\g \k}_{\g} 
+ C_{12} \p_{\x} \p^{\x} \p_{\k} \phi^{\t \g \m} \p_{\n} \p^{\n} \p^{\k} \phi_{\t \g \m} 
\\
{}&+ C_{13} \p_{\x} \p^{\x} \p_{\k} \phi^{\t \g \m} \p_{\n} \p^{\n} \p_{\g} \phi^{\k}_{\t  \m}
+ C_{14} \p_{\x} \p^{\x} \p_{\k} \phi^{\t \g \m} \p_{\t} \p_{\g} \p_{\m} \phi^{\n \k }_{\n }
+ C_{15}  \p_{\x} \p^{\x} \p_{\k} \phi^{\t \g \m} \p^{\k} \p_{\g} \p_{\m} \phi^{\n}_{\n \t }
\\
{}& + C_{16} \p_{\x} \p_{\k} \p_{\t} \phi^{\g \n}_{\g} \p^{\x} \p^{\k} \p^{\t} \phi^{\m}_{\m \n}
+ C_{17} \p_{\x} \p_{\k} \p_{\t} \phi^{\g \n}_{\g} \p_{\n} \p^{\x} \p^{\k} \phi^{\m \t}_{\m}
+  C_{18} \p_{\x} \p_{\t} \p_{\k} \phi^{\x \t}_{ \g} \p_{\m} \p_{\n} \p^{\k} \phi^{\m \n \g}
\\
{}&+  C_{19}  \p_{\x} \p_{\t} \p_{\k} \phi^{\x \t  \g} \p_{\m} \p_{\n} \p_{\g} \phi^{\m \n \k}
+  C_{20} \p_{\x} \p^{\x} \p_{\t} \phi^{\t \k}_{ \g} \p_{\n} \p^{\n} \p_{\m} \phi^{\m  \g}_{\k} 
+  C_{21}  \p_{\x} \p^{\x} \p_{\t} \phi^{\t \k  \g} \p_{\k} \p_{\g} \p_{\n} \phi^{\m  \n}_{\m}
\\
{}&+  C_{22}   \p_{\x} \p^{\x} \p_{\t} \phi^{\t \k  \g} \p_{\k} \p_{\m} \p_{\n} \phi^{\m  \n}_{\g} 
+  C_{23}  \p_{\x} \p^{\x} \p_{\t} \phi^{\t \k  \g} \p_{\k} \p_{\n} \p^{\n} \phi^{\m }_{\m \g}
+  C_{24}  \p_{\x} \p_{\t} \p_{\k} \phi^{\g \k }_{ \g} \p^{\x} \p^{\t} \p_{\n} \phi^{\m  \n}_{\m} 
\\
{}&+  C_{25}  \p_{\x} \p_{\t} \p_{\k} \phi^{\g \k }_{ \g} \p^{\x} \p_{\m} \p_{\n} \phi^{\m  \n \t}
+  C_{26}  \p_{\x} \p_{\t} \p_{\k} \phi^{\g \k }_{ \g} \p_{\n} \p^{\n} \p^{\x} \phi^{\m  \t}_{\m}
+  C_{27}  \p_{\x} \p^{\x} \p_{\t} \phi^{\g \k }_{ \g} \p^{\t} \p_{\m} \p_{\n} \phi^{\m  \n}_{\k}
\\
{}&+  C_{28} \p_{\x} \p^{\x} \p_{\t} \phi^{\g \k }_{ \g} \p_{\k} \p_{\m} \p_{\n} \phi^{\m  \n \t} 
+  C_{29}   \p_{\x} \p^{\x} \p_{\t} \phi^{\g \k }_{ \g} \p_{\n} \p^{\n} \p^{\t} \phi^{\m}_{\m \k}
+  C_{30} \p_{\x} \p^{\x} \p_{\t} \phi^{\g \k }_{ \g} \p_{\n} \p^{\n} \p_{\k} \phi^{\m \t}_{\m}
\\
{}& + C_{31} \p_{\x} \p_{\t} \p_{\k} \phi^{\x \t \k} \p_{\m} \p_{\n} \p_{\g} \phi^{\m \n \g}
+ C_{32} \p_{\x} \p_{\t} \p_{\k} \phi^{\x \t \k} \square \p^{\n} \phi^{\g}_{\g \n} 
+ C_{33}  \square \p_{\t} \phi^{\x \t}_{\x} \square \p^{\n} \phi^{\g}_{\g \n},
\end{split}
\end{equation}
where $\square = \partial_\alpha \partial^\alpha$. Note that, in the scalars multiplied by constants $C_1$ through $C_4$, $A$ and $B$ have 6 free indices, in the scalars multiplied by constants $C_{5}$ through $C_{17}$, $A$ and $B$ have 4 free indices, in the scalars multiplied by constants $C_{18}$ through $C_{30}$, $A$ and $B$ have 2 free indices, and in the scalars multiplied by constants $C_{31}$ through $C_{33}$, $A$ and $B$ have 0 free indices. This sorting is intentional: we will see that, as in the case of $N = M = 2$, these linear systems will decouple into factored curvature tensors of the `Riemann' and `Ricci' types. For clarity, we will treat these 4 types separately, as the $\mathcal{L}_{6}$, $\mathcal{L}_{4}$, $\mathcal{L}_{2}$ and $\mathcal{L}_{0}$ parts, respectively, of the general scalar, where the subscript refers to the number of free indices on $A$ and $B$ in the scalars of the given part. We can do this because the linear system of scalars identically decouples, with independent solutions for each of these four parts (there is no mixing between these four types of terms in the linear system of equations). Thus, for the above expression, we have $\mathcal{L} = \mathcal{L}_{6} + \mathcal{L}_{4} + \mathcal{L}_{2} + \mathcal{L}_{0}$.

Next we need to apply the gauge transformation $\phi_{\a \b \r}' = \phi_{\a \b \r} + \p_{\a} \lambda_{\b \r} + \p_{\b} \lambda_{\a \r} + \p_{\r} \lambda_{\a \b}$ to the general scalar and solve for the free coefficients such that the remaining expression is exactly gauge invariant.

\subsubsection{\boldmath Solving the $\mathcal{L}_{6}$ system of linear equations for spin-3 \\}

 Applying this transformation to $\mathcal{L}_{6}$ and combining like terms yields

\begin{equation}
\begin{split}
\mathcal{L}_{6} {}& = C_1 \p_{\x} \p_{\t} \p_{\k} \phi_{\g \m \n}\p^{\x} \p^{\t} \p^{\k} \phi^{\g \m \n} + C_2 \p_{\x} \p_{\t} \p_{\k}\phi_{\g \m \n}\p^{\x} \p^{\t} \p^{\g} \phi^{\k \m \n} + C_3 \p_{\x} \p_{\t} \p_{\k}\phi_{\g \m \n} \p^{\x} \p^{\m} \p^{\n} \phi^{\k \t \g}
\\
{}&+C_4 \p_{\x} \p_{\t} \p_{\k}\phi_{\g \m \n}\p^{\g} \p^{\m} \p^{\n} \phi^{\x \t \k}
+ \big(6C_1 + 2 C_2\big)\p_{\x} \p_{\t} \p_{\k} \p_{\g} \lambda_{\m \n} \p^{\x} \p^{\t} \p^{\k} \phi^{\g \m \n} 
\\
{}&+ \big(3 C_1 + C_2\big) \p_{\x} \p_{\t} \p_{\k} \p_{\g} \lambda_{\m \n} \p^{\x} \p^{\t} \p^{\k} \p^{\g} \lambda^{\m \n} 
+\big( 6C_1 +6 C_2 + 4C_3\big) \p_{\x} \p_{\t} \p_{\k} \p_{\g} \lambda_{\m \n} \p^{\x} \p^{\t} \p^{\k} \p^{\m} \lambda^{\g \n} 
\\
{}&+ \big(4 C_2 + 4C_3 \big) \p_{\x} \p_{\t} \p_{\k}\phi_{\g \m \n} \p^{\x} \p^{\t} \p^{\g} \p^{\m} \lambda^{\k \n} 
+ \big(2 C_2 + 5C_3 +9C_4\big) \p_{\x} \p_{\t} \p_{\k}\p_{\m} \lambda_{\g \m} \p^{\x} \p^{\t} \p^{\g} \p^{\m} \lambda^{\k \n} 
\\
{}&+ \big(2C_3 + 6C_4\big) \p_{\x} \p_{\t} \p_{\k}\phi_{\g \m \n} \p^{\x} \p^{\m} \p^{\n} \p^{\g} \lambda^{\k \t}.
\end{split}
 \end{equation}

The linear system of coefficients in front of the gauge parameter terms has the solution $C_1 = - C_4$, $C_2 = 3 C_4$, $C_3 = -3 C_4$ and $C_4 = C_4 = \tilde{A}$. Using this solution, the terms that depend on the gauge parameter all cancel and we are left with a gauge invariant expression. Remarkably, the remaining terms exactly factor into two independent 6 index tensors:

\begin{equation}
\begin{split}
\mathcal{L}_{6} = \tilde{A}{}& \big( \p_{\x} \p_{\t} \p_{\k}\phi_{\g \m \n} + \p_{\g} \p_{\t} \p_{\m}\phi_{\x \k \n} + \p_{\g} \p_{\n} \p_{\k}\phi_{\x \m \t}  + \p_{\x} \p_{\n} \p_{\m}\phi_{\g \k \t}
\\
{}& \qquad - \p_{\g} \p_{\n} \p_{\m}\phi_{\x \k \t} - \p_{\x} \p_{\t} \p_{\m}\phi_{\g \k \n} - \p_{\x} \p_{\n} \p_{\k}\phi_{\g \m \t} - \p_{\g} \p_{\t} \p_{\k}\phi_{\x \m \n} \big)
\\
{}& \times \big(\p^{\x} \p^{\t} \p^{\k} \phi^{\g \m \n} + \p^{\t} \p^{\m} \p^{\g} \phi^{\k \x \n} + \p^{\k} \p^{\n} \p^{\g} \phi^{\x \t \m} + \p^{\x} \p^{\m} \p^{\n} \phi^{\k \t \g}
\\
{}& \qquad - \p^{\g} \p^{\m} \p^{\n} \phi^{\x \t \k} - \p^{\x} \p^{\t} \p^{\m} \phi^{\k \n \g} - \p^{\k} \p^{\x} \p^{\n} \phi^{\t \m \g} - \p^{\t} \p^{\k} \p^{\g} \phi^{\x \m \n} \big).
\end{split}
\end{equation}

This is exactly the contraction of the spin-3 `Riemann' curvature tensor, $K^{\t \n \k \m \x \g}$, in equation (\ref{n3c})! Therefore we have derived the spin-3 'Riemann' curvature tensors by direct calculation. The contribution $\mathcal{L}_{6}$ has a unique gauge invariant solution, which is the contraction of the spin-3 curvature tensor: $\mathcal{L}_{6} = C_4 K_{\t \n \k \m \x \g} K^{\t \n \k \m \x \g}$. 

\subsubsection{\boldmath Solving the $\mathcal{L}_{4}$ system of linear equations for spin-3 \\}

Applying $\phi_{\a \b \r}' = \phi_{\a \b \r} + \p_{\a} \lambda_{\b \r} + \p_{\b} \lambda_{\a \r} + \p_{\r} \lambda_{\a \b}$ to $\mathcal{L}_{4}$ and combining like terms yields

\begin{equation}
\begin{split}
\mathcal{L}_4 {}& = C_5 \big( \p_{\x} \p_{\t } \p_{\k} \phi^{\x \m \n} \p_{\g} \p^{\k} \p^{\t} \phi^{\g}_{\m \n} \big) 
+ C_6 \big( \p_{\x} \p_{\t } \p_{\k} \phi^{\x \m \n} \p_{\m} \p_{\n} \p_{\g} \phi^{\g \k \t} \big) 
+ C_7 \big( \p_{\x} \p_{\t } \p_{\k} \phi^{\x \m \n} \p^{\k} \p_{\m} \p_{\g} \phi^{\g \t}_{ \n} \big) 
\\
{}&+ C_8 \big( \p_{\x} \p_{\t } \p_{\k} \phi^{\x \m \n} \p^{\k} \square \phi^{\t}_{\m \n} \big)
+ C_9 \big( \p_{\x} \p_{\t } \p_{\k} \phi^{\x \m \n} \p_{\n} \square \phi^{\k \t }_{\m} \big) 
+ C_{10} \big( \p_{\x} \p_{\t } \p_{\k} \phi^{\x \m \n} \p^{\t} \p^{\k} \p_{\m} \phi^{\g}_{\g \n} \big) 
\\
{}&+ C_{11} \big( \p_{\x} \p_{\t } \p_{\k} \phi^{\x \m \n} \p^{\t} \p_{\m} \p_{\n} \phi^{\g \k}_{\g} \big) 
+ C_{12} \big( \p_{\k} \square \phi^{\t \g \m} \p^{\k} \square \phi_{\t \g \m} \big) 
+ C_{13} \big( \p_{\k} \square \phi^{\t \g \m} \p_{\g} \square \phi^{\k}_{\t \m} \big) 
\\
{}&+ C_{14} \big( \p_{\k} \square \phi^{\t \g \m} \p_{\t} \p_{\g} \p_{\m} \phi^{\n \k }_{\n } \big) 
+ C_{15} \big( \p_{\k} \square \phi^{\t \g \m} \p^{\k} \p_{\g} \p_{\m} \phi^{\n}_{\n \t } \big) 
+ C_{16} \big( \p_{\x} \p_{\k} \p_{\t} \phi^{\g \n}_{\g} \p^{\x} \p^{\k} \p^{\t} \phi^{\m}_{\m \n} \big) 
\\
{}&+ C_{17} \big( \p_{\x} \p_{\k} \p_{\t} \phi^{\g \n}_{\g} \p_{\n} \p^{\x} \p^{\k} \phi^{\m \t}_{\m} \big) 
+\big(2C_5 + C_8 \big) \big( \p_{\x} \p_{\t } \p_{\k} \phi^{\x \m \n} \p^{\k} \p^{\t} \square \lambda_{\m \n} \big) 
\\
{}&+ \big(4C_5 + 2C_7 + 2C_{10}\big) \big( \p_{\x} \p_{\t } \p_{\k} \phi^{\x \m \n} \p_{\g} \p^{\k} \p^{\t} \p_{\m} \lambda^{\g}_{\n} \big) 
+ \big(2C_6 + C_9 \big) \big( \p_{\x} \p_{\t } \p_{\k} \phi^{\x \m \n} \p_{\m} \p_{\n} \square \lambda^{\k \t} \big)
\\
{}&+ \big(4C_6 + 2C_7 + 2C_{11} \big) \big( \p_{\x} \p_{\t } \p_{\k} \phi^{\x \m \n} \p_{\m} \p_{\n} \p_{\g} \p^{\k} \lambda^{\g \t} \big)
+ \big( 2C_7 + 2C_8 + 2C_9 \big) \big( \p_{\x} \p_{\t } \p_{\k} \phi^{\x \m \n} \p^{\k} \p_{\m} \square \lambda^{\t }_{\n} \big) 
 \\
{}&+ \big( C_8 + 6C_{12} + 2C_{13} \big) \big( \p_{\k} \square \phi^{\t \g \m} \p^{\k} \p_{\t} \square \lambda_{\g \m}\big) 
+ \big(2C_8 + C_9 + 2C_{15}\big) \big( \p_{\k} \square \phi^{\t \g \m} \p^{\k} \p_{\g} \p_{\m} \p^{\n} \lambda_{\t \n} \big) 
\\
{}&+ \big(C_9 + 4C_{13} \big) \big( \p_{\k} \square \phi^{\t \g \m} \p_{\g} \p_{\t} \square \lambda^{\k}_{\m} \big) 
+ \big( C_9 + 2C_{14} \big) \big( \p_{\k} \square \phi^{\t \g \m} \p_{\t} \p_{\g} \p_{\m} \p^{\n} \lambda^{\k}_{\n} \big)
\\
{}&+ \big( C_{10} + C_{11} \big) \big( \p_{\x} \p_{\t } \p_{\k} \phi^{\x \m \n} \p^{\t} \p_{\m} \p_{\n} \p^{\k} \lambda^{\g}_{\g} \big) 
+ \big(C_{10} + 2C_{15} \big) \big( \p_{\k} \p^{\g} \square \lambda^{\t \m} \p^{\k} \p_{\g} \p_{\m} \phi^{\n}_{\n \t } \big) 
\\
{}&+ \big( C_{10} + 2C_{11} + 4C_{17} \big) \big( \p_{\x} \p_{\k} \p_{\t} \phi^{\g \n}_{\g} \p_{\n} \p^{\x} \p^{\k} \p^{\m} \lambda^{\t}_{\m} \big) 
+ \big(C_{10} + 4C_{16} \big) \big( \p_{\x} \p_{\k} \p_{\t} \phi^{\g \n}_{\g} \p^{\x} \p^{\k} \p^{\t} \p^{\m} \lambda_{\n \m} \big) 
\\
{}&+ \big(C_{11} + 3C_{14} + C_{15}\big) \big( \p_{\k} \p^{\t} \square \lambda^{\g \m} \p_{\t} \p_{\g} \p_{\m} \phi^{\n \k }_{\n } \big) 
+ \big( C_{14} + C_{15} \big) \big( \p_{\k} \square \phi^{\t \g \m} \p_{\t} \p_{\g} \p_{\m} \p^{\k} \lambda^{\n}_{\n} \big) 
\\
{}&+ \big(2C_{16} + 2C_{17} \big) \big( \p_{\x} \p_{\k} \p_{\t} \phi^{\g \n}_{\g} \p_{\n} \p^{\x} \p^{\k} \p^{\t} \lambda^{\m}_{\m} \big)
+ \big( C_5 + C_8 + 3C_{12} + C_{13} \big) \big( \p_{\k} \p^{\g} \square \lambda^{\t \m} \p_{\g} \p^{\k} \square \lambda_{\t \m} \big) 
\\
{}&+ \big( 4C_5 + 2C_7 + 4C_8 + 2C_9 + 2C_{10} +4C_{15} \big) \big( \p_{\k} \p^{\g} \square \lambda^{\t \m} \p^{\k} \p_{\g} \p_{\m} \p^{\n} \lambda_{\t \n} \big)
\\
{}&+ \big( 2C_5 + C_7 + 2C_{10} + 4C_{16} \big) \big( \p_{\x} \p_{\k} \p_{\t} \p^{\g} \lambda^{\n}_{\g} \p^{\x} \p^{\k} \p^{\t} \p^{\m} \lambda_{\n \m} \big) 
\\
{}&+ \big( 2C_5 + 4C_6 + 3C_7 + 2C_{10} + 4C_{11} + 4C_{17} \big) \big( \p_{\x} \p_{\k} \p_{\t} \p^{\g} \lambda^{\n}_{\g} \p_{\n} \p^{\x} \p^{\k} \p^{\m} \lambda^{\t}_{\m} \big) 
\\
{}&+ \big( C_6 + C_9 + 2C_{13} \big) \big( \p_{\t } \p_{\k} \square \lambda^{\m \n} \p_{\m} \p_{\n} \square \lambda^{\k \t} \big) 
+ \big( C_{10} + C_{11} + 3C_{14} + 3C_{15} \big) \big( \p_{\k} \p^{\t} \square \lambda^{\g \m} \p_{\t} \p_{\g} \p_{\m} \p^{\k} \lambda^{\n}_{\n} \big) 
\\
{}&+ \big( 4C_6 + 2C_7 + 2C_8 + 4C_9 + 2C_{11} + 6C_{14} + 2C_{15} \big) \big( \p_{\k} \p^{\t} \square \lambda^{\g \m} \p^{\k} \p_{\g} \p_{\m} \p^{\n} \lambda_{\t \n} \big)
\\
{}&+ \big(C_7 + 2C_8 + 2C_9 + 6C_{12} + 6C_{13} \big) \big( \p_{\k} \p^{\t} \square \lambda^{\g \m} \p_{\g} \p^{\k} \square \lambda_{\t \m} \big) 
\\
{}&+ \big( 2C_{10} + 2C_{11} + 4C_{16} + 4C_{17} \big) \big( \p_{\x} \p_{\t } \p_{\k} \p^{\m} \lambda^{\x \n} \p^{\t} \p_{\m} \p_{\n} \p^{\k} \lambda^{\g}_{\g} \big) 
+ \big( C_{16} + C_{17} \big) \big( \p_{\x} \p_{\k} \p_{\t} \p^{\n} \lambda^{\g}_{\g} \p_{\n} \p^{\x} \p^{\k} \p^{\t} \lambda^{\m}_{\m} \big),
\end{split}
\end{equation}
which has the solution $C_{5} = -2C_{17}$, $C_{6} = 2C_{17}$, $C_{7} = 0$, $C_{8} = 4C_{17}$, $C_{9} = -4C_{17}$, $C_{10} = 4C_{17}$, $C_{11} = -4C_{17}$, $C_{12} = -C_{17}$, $C_{13} = C_{17}$, $C_{14} = 2C_{17}$, $C_{15} = -2C_{17}$, $C_{16} = -C_{17}$ and $C_{17} = C_{17} = \tilde{B}$. Using this solution, the terms that depend on the gauge parameter all cancel and we are left with a gauge invariant expression. Remarkably, the remaining terms exactly factor into two independent 4 index tensors:

\begin{equation}
\begin{split} 
\mathcal{L}_{4}  = {}&
\tilde{B} \big( \p^{\n} \square \phi^{\t \k \m}+ \p^{\m} \p^{\n} \p^{\t} \phi^{\g \k}_{\g}+\p_{\x} \p^{\t } \p^{\k} \phi^{\x \m \n} + \p_{\x} \p^{\m } \p^{\k} \phi^{\x \t \n}
\\
{}&\qquad -\p^{\k} \square \phi^{\t \n \m} - \p^{\m} \p^{\k} \p^{\t} \phi^{\g \n}_{\g} -\p_{\x} \p^{\t } \p^{\n} \phi^{\x \m \k} - \p_{\x} \p^{\m } \p^{\n} \phi^{\x \t \k} \big) 
\\
{}& \times \big( \p_{\n} \square \phi_{\t \k \m} + \p_{\n} \p_{\t} \p_{\m} \phi^{\g}_{\g \k} + \p_{\t } \p_{\k} \p^{\x} \phi_{\x \m \n} + \p_{\m } \p_{\k} \p^{\x}\phi_{\x \t \n} 
\\
{}& \qquad - \p_{\k} \square \phi_{\n \t \m} - \p_{\t} \p_{\k} \p_{\m} \phi^{\g }_{\g \n } - \p_{\m } \p_{\n} \p^{\x} \phi_{\x \t \k} - \p_{\t } \p_{\n} \p^{\x} \phi_{\x \m \k} \big).
\end{split}
\end{equation}

But this is exactly the contraction of the first spin-3 `Ricci' curvature tensor, $K^{\n \k \t \m}$! Therefore,  $\mathcal{L}_{4}$ has a unique gauge invariant solution, which is the contraction of the first spin-3 `Ricci' curvature tensor: $\mathcal{L}_{4} = \tilde{B} 4 K_{\n \k \t \m} K^{\n \k \t \m}$.

\subsubsection{\boldmath Solving the $\mathcal{L}_{2}$ system of linear equations for spin-3 \\}

Applying $\phi_{\a \b \r}' = \phi_{\a \b \r} + \p_{\a} \lambda_{\b \r} + \p_{\b} \lambda_{\a \r} + \p_{\r} \lambda_{\a \b}$ to $\mathcal{L}_{2}$ and combining like terms yields

\begin{equation}
\begin{split}
\mathcal{L}_{2} {}& =  C_{18} \big( \p_{\x} \p_{\t} \p_{\k} \phi^{\x \t}_{ \g} \p_{\m} \p_{\n} \p^{\k} \phi^{\m \n \g}\big) 
+ C_{19} \big(\p_{\x} \p_{\t} \p_{\k} \phi^{\x \t \g} \p_{\m} \p_{\n} \p_{\g} \phi^{\m \n \k} \big) 
+ C_{20} \big( \p_{\t} \square \phi^{\t \k}_{ \g} \p_{\m} \square \phi^{\m \g}_{\k} \big) 
\\
{}&+ C_{21} \big( \p_{\t} \square \phi^{\t \k \g} \p_{\k} \p_{\g} \p_{\n} \phi^{\m \n}_{\m} \big)
+ C_{22} \big( \p_{\t} \square \phi^{\t \k \g} \p_{\k} \p_{\m} \p_{\n} \phi^{\m \n}_{\g} \big) 
+ C_{23} \big( \p_{\t} \square \phi^{\t \k \g} \p_{\k} \square \phi^{\m }_{\m \g} \big) 
\\
{}&+ C_{24} \big( \p_{\x} \p_{\t} \p_{\k} \phi^{\g \k }_{ \g} \p^{\x} \p^{\t} \p_{\n} \phi^{\m \n}_{\m} \big) 
+ C_{25} \big( \p_{\x} \p_{\t} \p_{\k} \phi^{\g \k }_{ \g} \p^{\x} \p_{\m} \p_{\n} \phi^{\m \n \t} \big) 
+ C_{26} \big( \p_{\x} \p_{\t} \p_{\k} \phi^{\g \k }_{ \g} \square \p^{\x} \phi^{\m \t}_{\m} \big)
\\
{}&+ C_{27} \big( \square \p_{\t} \phi^{\g \k }_{ \g} \p^{\t} \p_{\m} \p_{\n} \phi^{\m \n}_{\k} \big) 
+ C_{28} \big( \square \p_{\t} \phi^{\g \k }_{ \g} \p_{\k} \p_{\m} \p_{\n} \phi^{\m \n \t} \big) 
+ C_{29} \big( \p_{\t} \square \phi^{\g \k }_{ \g} \p^{\t} \square \phi^{\m}_{\m \k} \big) 
\\
{}&+ C_{30} \big( \p_{\t} \square \phi^{\g \k }_{ \g} \p_{\k} \square \phi^{\m \t}_{\m} \big)
+ \big(4C_{18} + C_{22} + 2C_{27}\big) \big(\p_{\x} \p_{\t} \p_{\k} \phi^{\x \t}_{ \g} \p_{\n} \p^{\k} \square \lambda^{\n \g} \big) 
\\
{}&+ \big(2C_{18} + 2C_{19} +2C_{25}\big) \big( \p_{\x} \p_{\t} \p_{\k} \phi^{\x \t \g} \p_{\m} \p_{\n} \p_{\g} \p^{\k} \lambda^{\m \n} \big) 
\\
{}&+ \big(4C_{19} + C_{22} + 2C_{28}\big) \big( \p_{\x} \p_{\t} \p_{\k} \phi^{\x \t \g} \p_{\n} \p_{\g} \square \lambda^{\n \k} \big)
+ 2C_{20} \big( \p_{\t} \square \phi^{\t \k}_{ \g} \square \square \lambda^{\g}_{\k} \big) 
\\
{}&+ \big(4C_{20} + 2C_{22} + 2C_{23}\big) \big( \p_{\t} \square \phi^{\t \k}_{ \g} \p_{\m} \p^{\g} \square \lambda^{\m}_{\k} \big) 
+ \big(2C_{21} + C_{22} \big) \big(\p_{\t} \square \phi^{\t \k \g} \p_{\k} \p_{\g} \p_{\n} \p^{\m} \lambda^{\n}_{\m} \big) 
\\
{}&+ \big(C_{21} + C_{23}\big) \big( \p_{\t} \square \phi^{\t \k \g} \p_{\k} \p_{\g} \square \lambda^{\m}_{\m} \big)
+ \big(2C_{21} + 2C_{25} + 2C_{26}\big) \big(\p_{\t} \p^{\k} \square \lambda^{\t \g} \p_{\k} \p_{\g} \p_{\n} \phi^{\m \n}_{\m} \big) 
\\
{}&+ C_{21} \big( \square \square \lambda^{\k \g} \p_{\k} \p_{\g} \p_{\n} \phi^{\m \n}_{\m} \big) 
+ C_{22} \big( \square \square \lambda^{\k \g} \p_{\k} \p_{\m} \p_{\n} \phi^{\m \n}_{\g} \big) 
+ \big( C_{23} + 2C_{27} + 4C_{29} \big) \big( \p_{\t} \p^{\k} \square \lambda^{\t \g} \p_{\k} \square \phi^{\m }_{\m \g} \big)
\\
{}&+ C_{23} \big( \square \square \lambda^{\k \g} \p_{\k} \square \phi^{\m }_{\m \g} \big) 
+ \big(C_{23} + 2C_{28} + 4C_{30}\big) \big( \p_{\t} \square \phi^{\g \k }_{ \g} \p_{\k} \p^{\m} \square \lambda^{\t}_{ \m} \big) 
\\
{}&+ \big(4C_{24} + C_{25}\big) \big(\p_{\x} \p_{\t} \p_{\k} \phi^{\g \k }_{ \g} \p^{\x} \p^{\t} \p_{\n} \p^{\m} \lambda^{\n}_{ \m} \big) 
+ \big(C_{25} + C_{27} + C_{28} \big) \big( \p_{\t} \p^{\k} \square \lambda^{\g}_{\g} \p_{\k} \p_{\m} \p_{\n} \phi^{\m \n \t} \big) 
\\
{}&+ \big(2C_{24} + C_{26} \big) \big( \p_{\x} \p_{\t} \p_{\k} \phi^{\g \k }_{ \g} \p^{\x} \p^{\t} \square \lambda^{\m}_{\m} \big)
+ \big(2C_{26} + C_{27} + C_{28}\big) \big( \square \p_{\t} \phi^{\g \k }_{ \g} \p_{\k} \p_{\m} \p_{\n} \p^{\t} \lambda^{\m \n} \big) 
\\
{}&+ \big(C_{26} + 2C_{29} + 2C_{30}\big) \big( \p_{\t} \square \phi^{\g \k }_{ \g} \p_{\k} \p^{\t} \square \lambda^{\m}_{\m} \big)
+ \big(C_{24} + C_{26} + C_{29} + C_{30} \big) \big( \p_{\t} \p^{\k} \square \lambda^{\g}_{\g} \p_{\k} \p^{\t} \square \lambda^{\m}_{\m} \big)
\\
{}&+ \big(4C_{18} + 2C_{20} + 2C_{22} + 2C_{23} + 4C_{27} + 4C_{29} \big) \big( \p_{\t} \p_{\k} \square \lambda^{\t}_{\g} \p_{\n} \p^{\k} \square \lambda^{\n \g} \big) 
\\
{}&+ \big( 4C_{18} + 4C_{19} + 4C_{21} + 2C_{22} + 4C_{25} + 4C_{26} + 2C_{27} + 2C_{28} \big) \big( \p_{\t} \p_{\k} \square \lambda^{\t}_{\g} \p_{\m} \p_{\n} \p^{\k} \p^{\g} \lambda^{\m \n} \big)
\\
{}&+ \big(C_{18} + C_{19} + 4C_{24} +2C_{25} \big) \big( \p_{\x} \p_{\t} \p_{\k} \p^{\g} \lambda^{\k}_{ \g} \p^{\x} \p_{\m} \p_{\n} \p^{\t} \lambda^{\m \n} \big) 
\\
{}&+ \big( 4C_{19} + 2C_{20} + 2C_{22} + 2C_{23} + 4C_{28} + 4C_{30} \big) \big( \p_{\t} \p^{\g} \square \lambda^{\k}_{ \g} \p_{\k} \p^{\m} \square \lambda^{\t}_{ \m} \big)
\\
{}&+
\big( 4C_{20} + 2C_{22} + 2C_{23} \big) \big( \p_{\t} \p^{\k} \square \lambda^{\t}_{\g} \square \square \lambda^{\g}_{\k} \big) + C_{20} \big( \square \square \lambda^{\k}_{\g} \square \square \lambda^{\g}_{\k} \big) 
\\
{}&+ \big( 2C_{21} + 2C_{23} + 2C_{25} + 2C_{26} + 2C_{27} + 2C_{28} + 4C_{29} + 4C_{30} \big) \big( \p_{\t} \p^{\g} \square \lambda^{\k}_{ \g} \p_{\k} \p^{\t} \square \lambda^{\m}_{\m} \big)
\\
{}&+ \big(2C_{21} + C_{22}\big) \big( \square \square \lambda^{\k \g} \p_{\k} \p_{\g} \p_{\n} \p^{\m} \lambda^{\n}_{\m} \big) 
+ \big(C_{21} + C_{23} \big) \big( \square \square \lambda^{\k \g} \p_{\k} \p_{\g} \square \lambda^{\m}_{\m} \big) 
\\
{}&+ \big( 4C_{24} + C_{25} + 2C_{26} + C_{27} + C_{28} \big) \big( \p_{\t} \p^{\k} \square \lambda^{\g}_{\g} \p_{\k} \p_{\m} \p_{\n} \p^{\t} \lambda^{\m \n} \big)
,
\end{split}
\end{equation}
which has the solution $C_{18} = C_{18} = \tilde{C}$, $C_{19} = -\tilde{C}$, $C_{20} = 0$, $C_{21} = 0$, $C_{22} = 0$, $C_{23} = 0$, $C_{24} = 0$, $C_{25} = 0$, $C_{26} = 0$, $C_{27} = -2 \tilde{C}$, $C_{28} = 2 \tilde{C}$, $C_{29} = \tilde{C}$, and $C_{30} = -\tilde{C}$. Using this solution, the terms that depend on the gauge parameter all cancel and we are left with a gauge invariant expression. Remarkably, the remaining terms exactly factor into two independent 2 index tensors:

\begin{equation}
\begin{split}
\mathcal{L}_{2} = \tilde{C} 
{}&
\big( \p_{\m} \p_{\n} \p^{\k} \phi^{\m \n \g} 
- \p_{\m} \p_{\n} \p^{\g} \phi^{\m \n \k} 
+ \p^{\g} \square \phi^{\m \k }_{ \m} 
- \p^{\k} \square \phi^{\m \g }_{ \m} \big)
\\
{}& \times \big( \p_{\x} \p_{\t} \p_{\k} \phi^{\x \t}_{ \g} 
- \p_{\x} \p_{\t} \p_{\g} \phi^{\x \t}_{\k} 
- \square \p_{\k} \phi^{\x }_{ \x \g} 
+ \square \p_{\g} \phi^{\x }_{ \x \k} \big).
\end{split}
\end{equation}

But this is exactly the contraction of the second spin-3 `Ricci' curvature tensor, $K^{\k \g}$! Therefore, $\mathcal{L}_{2}$ has a unique gauge invariant solution, which is the contraction of the second spin-3 `Ricci' curvature tensor: $\mathcal{L}_{2} =  \tilde{C} K^{\k \g} K_{\k \g}$.

\subsubsection{\boldmath Solving the $\mathcal{L}_{0}$ system of linear equations for spin-3 \\}

Applying $\phi_{\a \b \r}' = \phi_{\a \b \r} + \p_{\a} \lambda_{\b \r} + \p_{\b} \lambda_{\a \r} + \p_{\r} \lambda_{\a \b}$ to $\mathcal{L}_{0}$ and combining like terms yields
\begin{equation}
\begin{split}
\mathcal{L}_{0} {}& = C_{31} \big( \p_{\x} \p_{\t} \p_{\k}\phi^{\x \t \k} \p_{\m} \p_{\n} \p_{\g} \phi^{\m \n \g} \big) 
+ C_{32} \big( \p_{\x} \p_{\t} \p_{\k} \phi^{\x \t \k} \square \p^{\n} \phi^{\g}_{\g \n} \big) 
\\
{}&+ C_{33} \big( \square \p_{\t} \phi^{\x \t}_{\x} \square \p^{\n} \phi^{\g}_{\g \n} \big) 
+\big(6C_{31} + 2C_{32}\big) \big( \p_{\x} \p_{\t} \p_{\k}\phi^{\x \t \k} \p_{\m} \p_{\n} \square \lambda^{\n \m} \big) 
\\
{}&+ C_{32} \big( \p_{\x} \p_{\t} \p_{\k} \phi^{\x \t \k} \square \square \lambda^{\g}_{\g} \big) 
+ \big(3 C_{32} + 4C_{33} \big) \big( \square \p_{\t} \phi^{\x \t}_{\x} \square \p^{\n} \p^{\g} \lambda_{\n \g} \big) 
\\
{}&+ 2C_{33} \big( \square \p_{\t} \phi^{\x \t}_{\x} \square \square \lambda^{\g}_{\g} \big)
+ \big(9C_{31} + 6C_{32} + 4C_{33} \big) \big( \p_{\t} \p_{\k}\square \lambda^{\t \k} \p_{\m} \p_{\n} \square \lambda^{\n \m} \big) 
\\
{}&+ \big(3C_{32} + 4C_{33} \big) \big(\square \p_{\t} \p^{\x} \lambda^{\t}_{ \x} \square \square \lambda^{\g}_{\g} \big) 
+ C_{33} \big( \square \square \lambda^{\x}_{\x} \square \square \lambda^{\g}_{\g} \big),
\end{split}
\end{equation}
which has the solution $C_{31} = C_{32} = C_{33} = 0$. Thus, $\mathcal{L}_{0} = 0$. This result is easily understood, since spin-$n$ models for $n = odd$ will have a scalar curvature tensor equal to zero, as in the case of electrodynamics, where $F = \eta_{\mu\nu} F^{\mu\nu} = 0$. 
Therefore, combining all parts, the Lagrangian density for the spin-3 case is 

\begin{equation}
\mathcal{L} = \tilde{A} K^{\t \n \k \m \x \g} K_{\t \n \k \m \x \g} + \tilde{B} K^{\t \n \k \x} K_{\t \n \k \x} + \tilde{C} K^{\n  \x} K_{\n  \x}.
\end{equation}

\subsection{\boldmath Scalars built from contracted spin-$n$ curvature tensors for $N = M = 4$}

For the $N = M = 4$ case, the most general Lagrangian density is the sum of all possible unique scalars of the form $\p \p \p \p \phi \p \p \p \p \phi$. This set of unique scalars is obtained by considering all the possible summation patterns that could occur in a scalar of the form $\p \p \p \p \phi \p \p \p \p \phi$. This is done using exactly the same method as for the $N=M=3$ case, except that, for $N = M = 4$, there are more possible summation patterns since there are 4 derivatives and 4 indices on $\phi$. Again we consider each scalar as a contraction of two terms which we call $A$ and $B$ and we group the possible scalars into categories based on the free indices of $A$ and $B$. In this case, it is possible for $A$ and $B$ to have 8, 6, 4, 2 or 0 free indices, so we have 5 categories. Next, as before, within each category, we consider all the possible contractions. By writing out a term for each possible summation pattern within each category, we obtain a comprehensive set of all the unique scalars of the form $\p \p \p \p \phi \p \p \p \p \phi$. 

 Again, the system of linear equations which is solved to find the values of the constant coefficients of $\mathcal{L}$ will decouple into independent linear systems based on the number of free indices (8, 6, 4, 2 or 0) on the $A$ and $B$ terms in the scalars the coefficients multiply. Therefore, we will again treat the most general Lagrangian density as the sum of the five types of Lagrangian densities; the most general Lagrangian density for the spin-4 case will then be of the form $\mathcal{L} = \mathcal{L}_{8} +  \mathcal{L}_{6} + \mathcal{L}_{4} + \mathcal{L}_{2} + \mathcal{L}_{0}$. For brevity (and because performing this calculation entirely by hand is a bit crazy), we will only directly solve for $\mathcal{L}_{8}$, since we know from the $N = M = 2$ and $N = M = 3$ cases that the remaining terms are built from contractions of the `Riemann' curvature tensor. The most general representation of $\mathcal{L}_8$ is

\begin{equation}
\begin{split}
\mathcal{L}_{8} {}& =  D_1 \p_{ \a}\p_{\g }\p_{ \m}\p_{ \n} \phi_{ \b \x \k \t} \p^{ \a}\p^{\g }\p^{ \m}\p^{ \n} \phi^{ \b \x \k \t} 
+ D_2 \p_{ \a}\p_{\g }\p_{ \m}\p_{ \n} \phi_{ \b \x \k \t} \p^{ \a}\p^{\g }\p^{ \m}\p^{ \t} \phi^{ \b \x \k \n}
\\
{}&+ D_3 \p_{ \a}\p_{\g }\p_{ \m}\p_{ \n} \phi_{ \b \x \k \t} \p^{ \a}\p^{\g }\p^{ \k }\p^{ \t} \phi^{ \b \x \m \n}
+ D_4 \p_{ \a}\p_{\g }\p_{ \m}\p_{ \n} \phi_{ \b \x \k \t} \p^{ \a}\p^{\x }\p^{ \k }\p^{ \t} \phi^{ \b \g \m \n}
\\
{}&+ D_5 \p_{ \a}\p_{\g }\p_{ \m}\p_{ \n} \phi_{ \b \x \k \t} \p^{ \b}\p^{\x }\p^{ \k}\p^{ \t} \phi^{ \a \g \m \n}.
\end{split}
\end{equation}

Next, we need to apply the gauge transformation $\phi_{ \b \x \k \t}'  = \phi_{ \b \x \k \t} +  \p_{\b} \lambda_{ \x \k \t} + \p_{\x }\lambda_{\b \k \t} + \p_{ \k}\lambda_{\b \x  \t} + \p_{\t} \lambda_{\b \x \k }$ to the general scalar and solve for the free coefficients such that the remaining expression is exactly gauge invariant. Applying this transformation to $\mathcal{L}_{8}$ and combining like terms yields

\begin{equation}
\hspace{-2cm}
\begin{split}
\mathcal{L}_{8} {}&= D_{1} \big( \p_{ \a}\p_{\g }\p_{ \m}\p_{ \n}\phi_{ \b \x \k \t} \p^{ \a}\p^{\g }\p^{ \m}\p^{ \n} \phi^{ \b \x \k \t} \big) 
+ D_{2} \big( \p_{ \a}\p_{\g }\p_{ \m}\p_{ \n} \phi_{ \b \x \k \t} \p^{ \a}\p^{\g }\p^{ \m}\p^{ \t} \phi^{ \b \x \k \n} \big) 
\\
{}&+ D_{3} \big( \p_{ \a}\p_{\g }\p_{ \m}\p_{ \n} \phi_{ \b \x \k \t}\p^{ \a}\p^{\g }\p^{ \k }\p^{ \t} \phi^{ \b \x \m \n} \big) 
+ D_{4} \big( \p_{ \a}\p_{\g }\p_{ \m}\p_{ \n} \phi_{ \b \x \k \t} \p^{ \a}\p^{\x }\p^{ \k }\p^{ \t} \phi^{ \b \g \m \n} \big) 
\\
{}&+ D_{5} \big( \p_{ \a}\p_{\g }\p_{ \m}\p_{ \n} \phi_{ \b \x \k \t} \p^{ \b}\p^{\x }\p^{ \k}\p^{ \t} \phi^{ \a \g \m \n} \big) 
+
\big(8D_{1} + 2D_{2}\big) \big( \p_{ \a}\p_{\g }\p_{ \m}\p_{ \n}\phi_{ \b \x \k \t} \p^{ \a}\p^{\g }\p^{ \m}\p^{ \n} \p^{\b} \lambda^{ \x \k \t} \big) 
\\
{}&+ \big(8D_{5} + 2D_{4} \big) \big( \p_{ \a}\p_{\g }\p_{ \m}\p_{ \n} \phi_{ \b \x \k \t} \p^{ \b}\p^{\x }\p^{ \k}\p^{ \t} \p^{\a} \lambda^{ \g \m \n} \big)
+ \big(6D_{4} + 4D_{3}\big) \big( \p_{ \a}\p_{\g }\p_{ \m}\p_{ \n} \phi_{ \b \x \k \t}\p^{ \a}\p^{\g }\p^{ \k }\p^{ \t} \p^{\b} \lambda^{ \x \m \n} \big) 
\\
{}&+ \big(4D_{3} + 6D_{2}\big) \big( \p_{ \a}\p_{\g }\p_{ \m}\p_{ \n} \phi_{ \b \x \k \t} \p^{ \a}\p^{\g }\p^{ \m}\p^{ \t} \p^{\b} \lambda^{ \x \k \n} \big)
+ \big(2D_{3} + 16D_{5} + 7D_{4}\big) \big( \p_{ \a}\p_{\g }\p_{ \m}\p_{ \n} \p_{\b} \lambda_{ \x \k \t}\p^{ \a}\p^{\g }\p^{ \k }\p^{ \t} \p^{\x }\lambda^{\b \m \n}\big) 
\\
{}&+ \big(9D_{4} + 10D_{3} + 6D_{2} \big) \big( \p_{ \a}\p_{\g }\p_{ \m}\p_{ \n} \p_{\x }\lambda_{\b \k \t} \p^{ \a}\p^{\x }\p^{ \k }\p^{ \t} \p^{\g }\lambda^{\b \m \n} 
\big) 
\\
{}&+ \big( 12D_{1} + 9D_{2} + 4D_{3}\big) \big( \p_{ \a}\p_{\g }\p_{ \m}\p_{ \n} \p_{\b} \lambda_{ \x \k \t} \p^{ \a}\p^{\g }\p^{ \m}\p^{ \n} \p^{\x }\lambda^{\b \k \t}
\big) 
+\big(4D_{1} + D_{2}\big) \big( \p_{ \a}\p_{\g }\p_{ \m}\p_{ \n} \p_{\t} \lambda_{\b \x \k } \p^{ \a}\p^{\g }\p^{ \m}\p^{ \t} \p^{\n} \lambda^{\b \x \k } \big).
\end{split}
\end{equation}

The linear system of coefficients has the solution $D_1 = D_1 = \tilde{D}$, $D_2 = -4 \tilde{D}$, $D_3 = 6 \tilde{D}$, $D_4 = -4 \tilde{D}$ and $D_5 = \tilde{D}$. Using this solution, the terms that depend on the gauge parameter all cancel and we are left with a gauge invariant expression. Remarkably, the remaining terms exactly factor into two independent 8 index tensors:

\begin{equation}
\begin{split}
\mathcal{L}_{8} {}&= \tilde{D} \Big( \p^{ \a}\p^{\g }\p^{ \m}\p^{ \n} \phi^{ \b \x \k \t} 
+ \p^{ \a}\p^{\g }\p^{ \k }\p^{ \t} \phi^{ \b \x \m \n} 
+ \p^{ \a}\p^{\m }\p^{ \x }\p^{ \t} \phi^{ \b \k \g \n}
+ \p^{ \a}\p^{\n }\p^{ \x }\p^{ \k} \phi^{ \b \t \g \m} 
\\
{}&+ \p^{ \g}\p^{\m }\p^{ \b }\p^{ \t} \phi^{ \x \k \a \n}
+ \p^{ \g}\p^{\n }\p^{ \b }\p^{ \k} \phi^{ \x \t \a \m} 
+ \p^{ \m}\p^{\n }\p^{ \b }\p^{ \x} \phi^{ \k \t \a \g}
+ \p^{ \b}\p^{\x }\p^{ \k}\p^{ \t} \phi^{ \a \g \m \n}
\\
{}&- 
\p^{ \a}\p^{\g }\p^{ \m}\p^{ \t} \phi^{ \b \x \k \n} 
- \p^{ \a}\p^{\g }\p^{ \n}\p^{ \k} \phi^{ \b \x \t \m} 
- \p^{ \a}\p^{\m }\p^{ \n}\p^{ \x} \phi^{ \b \k \t \g} 
 - \p^{ \g}\p^{\m }\p^{ \n}\p^{ \b} \phi^{ \x \k \t \a} 
 \\
{}&- \p^{ \a}\p^{\x }\p^{ \k }\p^{ \t} \phi^{ \b \g \m \n} 
- \p^{ \g}\p^{\b }\p^{ \k }\p^{ \t} \phi^{ \x \a \m \n} 
- \p^{ \m}\p^{\b }\p^{ \x }\p^{ \t} \phi^{ \k \a \g \n} 
- \p^{ \n}\p^{\b }\p^{ \x }\p^{ \k} \phi^{ \t \a \g \m} \Big) 
\\
{}& \times \Big( \p_{ \a}\p_{\g }\p_{ \m}\p_{ \n} \phi_{ \b \x \k \t} 
+ \p_{ \a}\p_{\g }\p_{ \k}\p_{ \t} \phi_{ \b \x \m \n} 
+ \p_{ \a}\p_{\x }\p_{ \m}\p_{ \t} \phi_{ \b \g \k \n} 
+ \p_{ \a}\p_{\x }\p_{ \k}\p_{ \n} \phi_{ \b \g \m \t} 
\\
{}&+ \p_{ \b}\p_{\g }\p_{ \m}\p_{ \t} \phi_{ \a \x \k \n}
+ \p_{ \b}\p_{\g }\p_{ \k}\p_{ \n} \phi_{ \a \x \m \t} 
+ \p_{ \b}\p_{\x }\p_{ \m}\p_{ \n} \phi_{ \a \g \k \t} 
+ \p_{ \b}\p_{\x }\p_{ \k}\p_{ \t} \phi_{ \a \g \m \n} 
\\
{}&- \p_{ \a}\p_{\g }\p_{ \m}\p_{ \t} \phi_{ \b \x \k \n} 
- \p_{ \a}\p_{\g }\p_{ \k}\p_{ \n} \phi_{ \b \x \m \t} 
- \p_{ \a}\p_{\x }\p_{ \m}\p_{ \n} \phi_{ \b \g \k \t}
- \p_{ \b}\p_{\g }\p_{ \m}\p_{ \n} \phi_{ \a \x \k \t} 
\\
{}&- \p_{ \a}\p_{\x }\p_{ \k}\p_{ \t} \phi_{ \b \g \m \n} 
- \p_{ \b}\p_{\g }\p_{ \k}\p_{ \t} \phi_{ \a \x \m \n} 
- \p_{ \b}\p_{\x }\p_{ \m}\p_{ \t} \phi_{ \a \g \k \n} 
- \p_{ \b}\p_{\x }\p_{ \k}\p_{ \n} \phi_{ \a \g \m \t} \Big).
\end{split}
\end{equation}

But this is exactly the contraction of the spin-4 `Riemann' curvature tensor, $K^{\a \b \g \x \m \k \n \t}$, in equation (\ref{n4c})! Therefore we have derived the spin-4 'Riemann' curvature tensor by direct calculation. The contribution $\mathcal{L}_{8}$ has a unique gauge invariant solution, which is the contraction of the spin-4 `Riemann' curvature tensor: $\mathcal{L}_{8} = \tilde{D} K^{\a \b \g \x \m \k \n \t} K_{\a \b \g \x \m \k \n \t}$.

In order to determine the possible form of the remaining spin-4 scalars, we will contract the curvature tensor in equation (\ref{n4c}); these contractions are known so we include them only for completeness. First, to find the rank 6 `Ricci' curvature tensor, we contract $\eta_{\a \b} K^{\a \g \b \x \m \k \n \t}$ to yield

\begin{equation}
\begin{split}
K^{\g \x \m \k \n \t}
{}&= \p^{ \m}\p^{ \n} \square \phi^{ \g \x \k \t} 
+ \p^{ \k }\p^{ \t} \square \phi^{ \g \x \m \n} 
+ \p_{ \a}\p^{\m }\p^{ \x }\p^{ \t} \phi^{ \g \k \a \n}
+ \p_{ \a}\p^{\n }\p^{ \x }\p^{ \k} \phi^{ \g \t \a \m} 
\\
{}&+ \p_{ \a}\p^{\m }\p^{ \g }\p^{ \t} \phi^{ \x \k \a \n}
+ \p_{ \a}\p^{\n }\p^{ \g }\p^{ \k} \phi^{ \x \t \a \m} 
+ \p^{ \m}\p^{\n }\p^{ \g }\p^{ \x} \phi^{ \a \k \t }_{\a}
+ \p^{ \g}\p^{\x }\p^{ \k}\p^{ \t} \phi^{ \a \m \n}_{\a}
\\
{}&- \p^{ \m}\p^{ \t} \square \phi^{ \g \x \k \n} 
- \p^{ \n}\p^{ \k} \square \phi^{ \g \x \t \m} 
- \p_{ \a}\p^{\m }\p^{ \n}\p^{ \x} \phi^{ \g \k \t \a} 
- \p_{ \a}\p^{\m }\p^{ \n}\p^{ \g} \phi^{ \x \k \t \a} 
\\
{}&- \p_{ \a}\p^{\x }\p^{ \k }\p^{ \t} \phi^{ \g \a \m \n} 
- \p_{ \a}\p^{\g }\p^{ \k }\p^{ \t} \phi^{ \x \a \m \n} 
- \p^{ \m}\p^{\g }\p^{ \x }\p^{ \t} \phi^{\a \k \n}_{\a} 
- \p^{ \n}\p^{\g }\p^{ \x }\p^{ \k} \phi^{ \a \t \m}_{\a}.
\end{split}
\end{equation}

This expression has the symmetries $K^{\g \x \m \k \n \t} = K^{\x \g \m \k \n \t} = K^{\g \x  \n \t \m \k} = - K^{\g \x \k \m \n \t}  = - K^{\g \x \m \k \t \n}$. There are three possible rank 4 contractions for the second `Ricci' curvature, although two are redundant (not independent).
The first (\ref{rank4Ricci1}) is found by contracting one of the indices in the symmetric pair and one in one of the antisymmetric pairs $\eta_{\k \m} K^{\m \x \k \g \n \t} $.
The second (\ref{rank4Ricci2}) is found by contracting one of the indices from each of the antisymmetric pairs $\eta_{\m \k} K^{\x \g \m \n \k \t}$.
The third that is equivalent to the first (\ref{rank4Ricci1}) is found by contracting both the indices of the symmetric pair $\eta_{\k \m}K^{\k \m \x \g \n \t}$. The rank 4 tensors are:

\begin{equation}
\begin{split}
\label{rank4Ricci1}
{K}^{\x \g \n \t} {}& =
2 \p^{ \x}\p^{ \n} \square \phi^{ \k \g \t}_{\k} 
+ 2\p^{ \g }\p^{ \t} \square \phi^{ \k \x \n}_{\k} + 2 \p_{ \k } \p_{ \a}\p^{\x }\p^{ \t} \phi^{ \k \g \a \n}
+ 2\p_{ \k } \p_{ \a}\p^{\n }\p^{ \g} \phi^{ \k \t \a \x} 
\\
{}&- 2\p^{ \x}\p^{ \t} \square \phi^{ \k \g \n}_{\k} - 2\p^{ \n}\p^{ \g} \square \phi^{ \k \t \x} _{\k} 
- 2 \p_{ \k } \p_{ \a}\p^{\x }\p^{ \n} \phi^{ \k \g \t \a} 
- 2\p_{ \k } \p_{ \a} \p^{ \g }\p^{ \t} \phi^{ \k \a \x \n} ,
\end{split}
\end{equation}

\begin{equation}
\begin{split}
\label{rank4Ricci2}
\hat{K} ^{\x \g \n \t} {}&= \square \square \phi^{ \x \g \n \t} + \p^{ \n }\p^{ \t} \square \phi^{ \k \x \g }_{\k} + \p_{\k } \p_{ \a}\p^{ \g }\p^{ \t} \phi^{ \x \n \a \k}
+ \p_{ \a}\p_{\k }\p^{ \g }\p^{ \n} \phi^{ \x \t \a \k} 
\\
{}&+ \p_{ \a}\p_{\k }\p^{ \x }\p^{ \t} \phi^{ \g \n \a \k}
+ \p_{ \a}\p_{\k }\p^{ \x }\p^{ \n} \phi^{ \g \t \a \k} + \p^{ \x }\p^{ \g} \square \phi^{ \a \n \t }_{\a}
+ \p^{ \x}\p^{\g }\p^{ \n}\p^{ \t} \phi^{ \a \k}_{\a \k}
\\
{}&- \p_{ \k}\p^{ \t} \square \phi^{ \x \g \n \k} - \p_{ \k}\p^{ \n} \square \phi^{ \x \g \t \k} 
- \p_{ \a}\p^{ \g} \square \phi^{ \x \n \t \a} - \p_{ \a}\p^{ \x} \square \phi^{ \g \n \t \a} 
\\
{}&- \p_{ \a}\p^{\g }\p^{ \n }\p^{ \t} \phi^{ \k \x \a }_{\k} - \p_{ \a}\p^{\x }\p^{ \n }\p^{ \t} \phi^{ \k \g \a} _{\k}
- \p_{ \k}\p^{\x }\p^{ \g }\p^{ \t} \phi^{\a \n \k}_{\a} - \p_{ \k}\p^{\x }\p^{ \g }\p^{ \n} \phi^{ \a \t \k}_{\a} ,
\end{split}
\end{equation}

where 
${K}^{\x \g \n \t}$ in (\ref{rank4Ricci1}) has symmetries and anti-symmetries $\bar{K}^{\x \g \n \t} = {K}^{ \n \t \x \g } = - {K}^{\g \x \n \t}  = - {K}^{\x \g \t \n}$
and $\hat{K} ^{\x \g \n \t}$ in (\ref{rank4Ricci2}) has symmetries $\hat{K} ^{\x \g \n \t} = \hat{K} ^{\g \x \n \t} = \hat{K} ^{\x \g \t \n} = \hat{K}^{\n \t \x \g}$. Contracting either of $\eta_{\x \g} \hat{K} ^{\x \g \n \t} $ (\ref{rank4Ricci2}) or $\eta_{\g \x} \bar{K}^{\x \n \g \t}$ (\ref{rank4Ricci1}) yields the same, unique, rank 2 tensor,

\begin{equation}
\begin{split}
K^{\n \t}
{}&= 2 \square \square \phi^{ \g \n \t} _{\g} 
+ 2 \p^{ \n }\p^{ \t} \square \phi^{ \k \g}_{ \k \g} + 2\p_{\k } \p_{ \a}\p_{ \g }\p^{ \t} \phi^{ \g \n \a \k}
+ 2 \p_{ \a}\p_{\k }\p_{ \g }\p^{ \n} \phi^{ \g \t \a \k} 
\\
{}&- 2 \p_{ \k}\p^{ \t} \square \phi^{ \g \n \k}_{\g} - 2 \p_{ \k}\p^{ \n} \square \phi^{ \g \t \k} _{\g} - 2 \p_{ \a}\p_{ \g} \square \phi^{ \g \n \t \a} - 2 \p_{ \a}\p_{\g }\p^{ \n }\p^{ \t} \phi^{ \k \g \a }_{\k},
\end{split}
\end{equation}
with symmetry $K^{\n \t} = K^{\t \n}$. As in the case of the standard (spin-2 curvature) Riemann tensor, we can derive a nonzero scalar curvature from this by contracting $ \eta_{\t \n} K^{\n \t}$, which yields

\begin{equation}
K = 2 \square \square \phi^{ \a \t }_{\a \t} + 2\p_{\k } \p_{ \a}\p_{ \g }\p_{ \t} \phi^{ \k \t \a \g} 
- 4 \p_{\k } \p_{ \a} \square \phi^{\t \k \a}_{\t}.      
 \end{equation}

Therefore, in the case of spin-4, we have a Lagrangian of the form $\mathcal{L} = \mathcal{L}_{8} +  \mathcal{L}_{6} + \mathcal{L}_{4} + \mathcal{L}_{2} + \mathcal{L}_{0}$, where the $\mathcal{L}_{4}$ contribution has two different possible contractions. The most general Lagrangian possible for spin-4 is then $\mathcal{L} = \tilde{D} K^{\a \b \g \x \m \k \n \t} K_{\a \b \g \x \m \k \n \t} + \tilde{E} K^{\g \x \m \k \n \t} K_{\g \x \m \k \n \t} + \Sigma_j  \tilde{F}_j {\bf{K}}^{\x \g \n \t} {\bf{K}}_{\x \g \n \t} + \tilde{G} K^{\n \t} K_{\n \t} + \tilde{H} K^2$, where $\Sigma_j  \tilde{F}_j {\bf{K}}^{\x \g \n \t} {\bf{K}}_{\x \g \n \t}$ represents all possible scalars built from the rank 4 curvature tensors. The 2 different curvature tensors $K^{\x \g \n \t}$ and $\hat{K} ^{\x \g \n \t}$  present an ambiguity problem not observed in the lower spin-$n$ models. We investigated this ambiguity by considering the most general scalar for rank 4 curvature tensors. The scalars built from each of these 2 curvature tensors are indeed solutions to the resulting linear system. Therefore we will not attempt to select one of these expressions as being superior to the others. This result shows that the curvature scalars for higher spin models can have more than one combination at each rank of curvature tensor.

\subsection{\boldmath Scalars built from contracted spin-$n$ curvature tensors for $N = M = n$}

We have shown by direct calculation that the curvature tensors of higher-spin gauge theories \cite{sorokin2005,bekaert2006} can be derived from the procedure in \cite{baker2019} without any a priori knowledge of their existence:

\begin{equation}
\begin{split}
F^{[\mu\nu]} ={}& \partial^\mu A^\nu - \partial^\nu A^\mu  ,
\label{n1c}
\end{split}
\end{equation}

\begin{equation}
\begin{split}
R^{[\mu\nu][\alpha\beta]} ={}&  \partial^\mu \partial^\beta h^{\nu\alpha} + \partial^\nu \partial^\alpha h^{\mu\beta} -\partial^\mu \partial^\alpha h^{\nu\beta} - \partial^\nu \partial^\beta h^{\mu\alpha}  ,
\label{n2c}
\end{split}
\end{equation}

\begin{equation}
\begin{split}
\label{n3c}
K^{[\t \n][\k \m] [\x \g]} {}&=  \p^{\x} \p^{\t} \p^{\k} \phi^{\g \m \n}  +   \p^{\t} \p^{\m} \p^{\g} \phi^{\k \x \n} +   \p^{\k} \p^{\n} \p^{\g} \phi^{\x \t \m} +  \p^{\x} \p^{\m} \p^{\n} \phi^{\k \t \g} 
\\
& - \p^{\g} \p^{\m} \p^{\n} \phi^{\x \t \k} - \p^{\x} \p^{\t} \p^{\m} \phi^{\k \n \g} 
 - \p^{\k} \p^{\x} \p^{\n} \phi^{\t \m \g}  - \p^{\t} \p^{\k} \p^{\g} \phi^{\x \m \n}  ,
\end{split}
\end{equation}

\begin{equation}
\begin{split}
\label{n4c}
K^{[\a \b] [\g \x] [\m \k] [\n \t]} {}&=  \p^{ \a}\p^{\g }\p^{ \m}\p^{ \n} \phi^{ \b \x \k \t}   
+	 \p^{ \a}\p^{\g }\p^{ \k }\p^{ \t}  \phi^{ \b \x \m \n}  	
+	 \p^{ \a}\p^{\m }\p^{ \x }\p^{ \t}  \phi^{ \b \k \g \n}	
 +	 \p^{ \a}\p^{\n }\p^{ \x }\p^{ \k}  \phi^{ \b \t \g \m}
\\
  &
+ 	  \p^{ \g}\p^{\m }\p^{ \b }\p^{ \t}  \phi^{ \x \k \a \n}
 + 	  \p^{ \g}\p^{\n }\p^{ \b }\p^{ \k}  \phi^{ \x \t \a \m} 
 + 	  \p^{ \m}\p^{\n }\p^{ \b }\p^{ \x}  \phi^{ \k \t \a \g}
    +     \p^{ \b}\p^{\x }\p^{ \k}\p^{ \t}  \phi^{ \a \g \m \n} 
\\
&
-    \p^{ \a}\p^{\g }\p^{ \m}\p^{ \t}   \phi^{ \b \x \k \n}   
-		 \p^{ \a}\p^{\g }\p^{ \n}\p^{ \k}   \phi^{ \b \x \t \m}  
- 	 \p^{ \a}\p^{\m }\p^{ \n}\p^{ \x}   \phi^{ \b \k \t \g}   
 -	 \p^{ \g}\p^{\m }\p^{ \n}\p^{ \b}   \phi^{ \x \k \t \a}   
\\   
 &-    	 \p^{ \a}\p^{\x }\p^{ \k }\p^{ \t}  \phi^{ \b \g \m \n}   
 -  	 \p^{ \g}\p^{\b }\p^{ \k }\p^{ \t}  \phi^{ \x \a \m \n}  
 -	 \p^{ \m}\p^{\b }\p^{ \x }\p^{ \t}  \phi^{ \k \a \g \n} 	 
-	 \p^{ \n}\p^{\b }\p^{ \x }\p^{ \k}  \phi^{ \t \a \g \m}  ,
\end{split}
\end{equation}

where the curvatures $K$ have $n$ pairs of antisymmetric indices for each spin-$n$ that are symmetric under interchange, and the tensor potentials $\phi$ are all totally symmetric. Note that equations (\ref{n1c}) and (\ref{n2c}), the electrodynamic field strength and linearized Riemann tensor, were derived using this procedure in \cite{baker2019}. 

Since we cannot compute the Lagrangian densities built from curvature tensors of all spin-$n$ models from the $N = M = n$ procedure, we can at best give some conjectures as to the expected form of these Lagrangian densities, based on the $n = 1$ through $n = 4$ cases. What we know is that the procedure is well defined for all $n$ for which the gauge transformations are generalized by equations (\ref{spin1trans}) to (\ref{spin4trans}). Since $n$ corresponds to the number of derivatives and the number of indices on the tensor potential, we can generalize $\mathcal{L}$ at each $n$ using the notation $\partial_{\mu_n} = \partial_\mu \dots \partial_\alpha$, which is a product of $n$ partial derivatives, and $\phi_{\nu_n} = \phi_{\nu \dots \beta}$, which is a completely symmetric tensor potential with $n$ indices. 
Then the general $\mathcal{L}$ will be the sum of all possible $i$ unique scalars of the form $(\partial_{\mu_n} \phi_{\nu_n})^2$, which we write as $\mathcal{L} = \Sigma_i C_i (\partial_{\mu_n} \phi_{\nu_n} )^2$. For this generalization, we can make the following two conjectures: (i) Under a spin-$n$ gauge transformation, a higher-spin Lagrangian density of the form $\mathcal{L} = \Sigma_i C_i (\partial_{\mu_n} \phi_{\nu_n} )^2$ will have a unique gauge invariant solution that decouples into contractions of the spin-$n$ 'Riemann' curvature tensor and its `Ricci' tensors and scalar. This decoupling will be of the form $\mathcal{L} = \mathcal{L}_{2n} + \dots + \mathcal{L}_{4} +\mathcal{L}_{2} + \mathcal{L}_{0}$, where $\mathcal{L}_{2n}$ is the contracted curvature tensors for the spin-$n$ theory. Each of these curvature tensors will have $n$ pairs of antisymmetric indices and the pairs will all be symmetric with one another. (ii) For $n = odd$, the term $\mathcal{L}_{0} = 0$, leaving no `Ricci' scalar in such models, as seen in electrodynamics and spin-3. We emphasize that in order to have an exactly gauge invariant Lagrangian---built from quadratic combinations of derivatives of potentials---for a higher spin model, one uniquely requires the contraction of independently gauge invariant `field strength' tensors, known as the curvature tensors of higher spin gauge theories.

\section{Conclusions}

The curvature tensors of higher-spin gauge theories have been derived from first principles; that is, without any a priori knowledge of their existence. Using a procedure that considers the most general linear combination of scalars built from quadratic combinations of $N$ order of derivatives and $M$ rank of tensor potential, we explored the general case of $N = M = n$, under the spin-$n$ gauge transformations. It had been shown in \cite{baker2019} that the $N = M = 1$ case uniquely determines the contraction of the field strength tensor of electrodynamics $\mathcal{L} = C F_{\mu\nu} F^{\mu\nu}$ and the $N = M = 2$ case uniquely determines the contraction of the linearized `Riemann' and `Ricci' tensors $\mathcal{L} = \tilde{a} R_{\mu\nu\alpha\beta} R^{\mu\nu\alpha\beta} + \tilde{b} R_{\mu\nu} R^{\mu\nu} + \tilde{c} R^2$. In this article we first considered the $N = M = 3$ case, under the spin-3 gauge transformation. As expected, based on the previous result, this system of linear equations decoupled into unique solutions that correspond to the contraction of the well-known curvature tensor for spin-3 and its `Ricci' forms $\mathcal{L} = \tilde{A} K^{\t \n \k \m \x \g} K_{\t \n \k \m \x \g} + \tilde{B} K^{\t \n \k \x} K_{\t \n \k \x} + \tilde{C} K^{\n  \x} K_{\n  \x}$. 
This is a notable result for two reasons: (i) these curvature tensors were uniquely derived without any knowledge of their existence influencing the procedure and 
(ii) it provides a method for explicitly deducing these expressions, which are typically just generalized from lower rank models using inductive arguments and known symmetries (by taking the curl on each index of a totally symmetric rank-$n$ field for each spin-$n$ \cite{damour1987}). 
The same process was then considered for the $N = M = 4$ case, for the highest rank only, and, again, contraction of the spin-4 curvature tensor $\mathcal{L}_{8} = \tilde{D} K^{\a \b \g \x \m \k \n \t} K_{\a \b \g \x \m \k \n \t}$
 was uniquely determined. By considering contractions of this tensor, we wrote down the possible form of the most general Lagrangian for the $N = M = 4$ case. 
 Finally, we provided some conjectures regarding the $N = M = n$ case. 

 What is interesting to note is that in \cite{baker2019} it was shown that, since the $N=M=n$ Lagrangians are exactly gauge invariant, one can use the Bessel-Hagen result from Noether's first theorem \cite{besselhagen1921,noether1918} to derive gauge invariant energy-momentum tensors by fixing the remaining free coefficients in the Lagrangians. For the $N = M = 1$ case, this procedure results in the coefficients being fixed such that the Lagrangian obtained is the Lagrangian of classical electrodynamics while, for the $N=M=2$ case, it results in the coefficients being fixed such that the Lagrangian obtained is that of the linearized Gauss-Bonnet gravity model. In this article, we obtained the result that, for higher spin models $N=M \ge 3$, if one imposes the requirement of having exactly gauge invariant Lagrangians (not merely invariant up to a surface term) built from quadratic combinations, then one again uniquely obtains the contraction of independently gauge invariant curvature tensors as a direct result of this requirement. In the past, such Lagrangians have been postulated in the physics literature but never derived \cite{francia2010}. We again acknowledge that the higher derivative models $N=M \ge 3$ have problems with unitary and renormalizability \cite{abe2019}, and have no obvious predicative utility (building the models associated to these Lagrangian densities is not the purpose of our article, our purpose is to derive the the curvature tensors of higher spin gauge theories without a priori knowledge of their existence). However, having more complicated exactly gauge invariant actions can provide useful toy models to answer questions about the generalization of the Noether and Bessel-Hagen methods to more complicated theories, such as the recent use of both the $N=M=1$ case (electrodynamics) and the $N=M=2$ case (linearized Gauss-Bonnet gravity) in disproving the notion of general equivalence between the Noether and Hilbert energy-momentum tensors \cite{baker2021}. In addition the Noether identities can be used to generalize beyond the free field consideration \cite{kiriushcheva2014}. Whether applying the Noether/Bessel-Hagen method to $N=M \ge 3$ Lagrangians will uniquely fix the free coefficients of these Lagrangians such that gauge invariant energy-momentum tensors are derived, as for $N=M=1$ and $N=M=2$, is the subject of future work.

\begin{acknowledgments}
The authors are grateful to L. Bruce-Robertson, N. Kiriushcheva, S. Kuzmin and D.G.C. McKeon for numerous discussions and suggestions during the preparation of this article.
 \end{acknowledgments}

\bibliography{HigherSpinReferencesV2}{}

\end{document}